\documentclass[aps,pra,twocolumn,showpacs]{revtex4-1}

\usepackage{graphicx}
\usepackage{amsmath,amssymb,amsfonts}
\usepackage{natbib}

\newcommand{\bd}{\begin{displaymath}}
\newcommand{\ed}{\end{displaymath}}
\newcommand{\be}{\begin{equation}}
\newcommand{\ee}{\end{equation}}
\newcommand{\ba}{\begin{eqnarray}}
\newcommand{\ea}{\end{eqnarray}}

\begin{document}

\title[Bohm's approach to quantum mechanics]{Bohm's approach to quantum mechanics:\\
Alternative theory or practical picture?}

\author{A. S. Sanz}
\email{a.s.sanz@fis.ucm.es}
\affiliation{Department of Optics, Faculty of Physical Sciences,
Universidad Complutense de Madrid,\\
Pza.\ Ciencias 1, Ciudad Universitaria E-28040 Madrid, Spain}

\begin{abstract}
Since its inception Bohmian mechanics has been generally regarded
as a hidden-variable theory aimed at providing an {\it objective}
description of quantum phenomena.
To date, this rather narrow conception of Bohm's proposal has caused it
more rejection than acceptance.
Now, after 65 years of Bohmian mechanics, should still be such an
interpretational aspect the prevailing appraisal?
Why not favoring a more pragmatic view, as a legitimate picture of
quantum mechanics, on equal footing in all respects with any other more
conventional quantum picture?
These questions are used here to introduce a discussion on an
alternative way to deal with Bohmian mechanics at present, enhancing
its aspect as an efficient and useful picture or formulation to tackle,
explore, describe and explain quantum phenomena where phase and
correlation ({\it entanglement}) are key elements.
This discussion is presented through two complementary blocks.
The first block is aimed at briefly revisiting the historical context
that gave rise to the appearance of Bohmian mechanics, and how this
approach or analogous ones have been used in different physical
contexts.
This discussion is used to emphasize a more pragmatic view to the
detriment of the more conventional hidden-variable (ontological)
approach that has been a leitmotif within the quantum foundations.
The second block focuses on some particular formal aspects of Bohmian
mechanics supporting the view presented here, with special emphasis on
the physical meaning of the local phase field and the associated
velocity field encoded within the wave function.
As an illustration, a simple model of Young's two-slit experiment
is considered.
The simplicity of this model allows to understand in an easy manner
how the information conveyed by the Bohmian formulation relates to
other more conventional concepts in quantum mechanics.
This sort of pedagogical application is also aimed at showing the
potential interest to introduce Bohmian mechanics in undergraduate
quantum mechanics courses as a working tool rather than merely an
alternative interpretation.
\end{abstract}

\pacs{03.65.Ca,03.65.Ge,03.65.Ta,03.75.-b}


\maketitle


\section{Introduction}
\label{sec1}

In 1917 Einstein settled down the theoretical foundations for the maser
and the laser with the introducing of the notion of stimulated radiation
emission \cite{einstein:PhysZeits:1917}.
By the end of the same year, David Bohm, Einstein's ``spiritual son'',
was born in Wilkes-Barre, Pennsylvania.
About 35 years later, Bohm proposed what is now widely known as Bohmian
mechanics \cite{bohm:PR:1952-1,bohm:PR:1952-2}, an alternative way to
look at quantum phenomena that would have a strong influence on John
Bell and his celebrated theorem -- which led to new research areas in
quantum  mechanics, such as the quantum information theory and the
so-called quantum technologies.
But, just when this work on the reformulation of quantum mechanics has
turned 65, how much of its pedagogical value to describe, understand
and explain quantum phenomena is actually known out of the quantum
foundations community?
Indeed, one may even wonder whether it would not be worth taking it out
of such a community, and incorporate it into the curricular scopes of
quantum mechanics courses, in the same way we learn, for instance,
to operate with the Schr\"odinger and Heisenberg formulations.

Since the 1930s there has been an active discussion around the
question of whether it is possible to find a compromise between the
internal consistency of the quantum theory and its apparently
statistical or probabilistic nature, quite distant from the classical
causality we are used to.
In 1952, moved by the need to understand and provide a clear answer to
this fundamental question, Bohm proposed \cite{bohm:PR:1952-1,bohm:PR:1952-2}
what he called an ``interpretation of the quantum theory in terms of
``hidden'' variables''.
Several decades later, this interpretation started being referred to as
{\it Bohmian mechanics} and became one of the warhorse issues
within the still open debate on the interpretation of the quantum
realm, with a strong influence on the perception and understanding of
Reality and the Universe we live in \cite{bohm-bk:1980,hiley-chap:2016}.

Leaving aside such deep matters and remaining at a bare formal level --
i.e., taking on a more pragmatic viewpoint -- what Bohmian mechanics
does is to provide us with a precise mathematical language to tackle
quantum
problems \cite{wyatt-bk,chattaraj-bk,hughes-bk,sanz-bk-2,oriols-bk,benseny:EPJD:2014},
complementary to the mathematical language involved in other more
widely known quantum
pictures \cite{note1},
such as those proposed by Schr\"odinger, Heisenberg, Wigner and Moyal, or Feynman,
for instance (for the interested reader, an enlightening perspective on the different
formulations of quantum mechanics is provided by Styer {\it et al.}\ in ``Nine
formulations of quantum mechanics'' \cite{styer:AJP:2002}).
Such a language has a strong reminiscence of classical
hydrodynamics, offering an alternative way to represent quantum
phenomena in terms of swarms of time-evolving streamlines or
trajectories. In spite of this neat single-event dynamics, the
results obtained from a statistical analysis of the behavior of
these swarms of trajectories is in direct correspondence with the
expectation value associated with quantum observables in those other
quantum pictures.

The main goal of this work is to introduce a discussion on an updated
view of Bohmian mechanics beyond its more widely known ontological
side.
This is done taking advantage of the gradual transition in the
perception of this approach, from its origins as a nonlocal
hidden-variable model to the current wider view as a quantum picture.
This discussion is thus guided by two intertwined questions:
\begin{itemize}
 \item[$\bullet$] After 65 years of Bohmian mechanics, should its
 interpretational aspect still be its prevailing appraisal?
 \item[$\bullet$] Is Bohmian mechanics a legitimate picture of quantum
 mechanics, on equal footing in all respects with any other quantum
 picture?
\end{itemize}
Accordingly, the work has been organized around two complementary
building blocks.
The first block, Sec.~\ref{sec2}, enters a discussion on the origin and
role of Bohmian mechanics as a hidden-variable model \cite{note2}, introducing
the issue through two of its most common criticisms and concluding it
with a discussion on how the idea of monitoring waves (not necessarily
quantum ones) by means of streamlines or trajectories has been used in
the literature with other purposes than fundamental ones.
Because it covers both fundamental and computational aspects, this
block offers a wider perspective on questions that do not usually
appear together in the context of the quantum foundations in spite of
the relevance of a joint view.
The second block, Sec.~\ref{sec3}, addresses in a simplified manner
the more basic formal aspects of Bohmian mechanics, useful to interpret
quantum outcomes obtained from Schr\"odinger's equation on a
hydrodynamic basis with the aid of Bohmian trajectories.
Other Bohmian-related quantities, such as the local quantum phase field,
the quantum velocity, or the quantum potential are also discussed.
As an illustration of these concepts, and particularly with the
purpose to explicitly show how Bohmian mechanics operates or what
kind of information it may provide us, this block also includes a revision
and discussion of a simple implementation of Young's two-slit experiment.
Finally, in Sec.~\ref{sec5} a series of remarks summarize the main points
stressed in this work.


\section{The general ``picture''}
\label{sec2}


\subsection{What's wrong with Bohmian mechanics?}
\label{sec21}

Different criticisms can be risen against Bohmian
mechanics \cite{holland-bk}.
Among them, there are two that are particularly relevant to the
discussion in this work.
The first criticism is very common: because the outputs rendered
by Bohmian mechanics are equivalent to those provided by other quantum
pictures, this approach is redundant and, therefore, unnecessary.
Trivial, but acceptable.
Although it is also unfair, unless the same
is rigorously applied to other pictures.
Of course, this leads us to the absurd situation that all pictures except one are
also redundant and unnecessary, which is not the case.
The Schr\"odinger and Heisenberg pictures, for instance, are
conceptually different, not only in their origin, but also concerning
their mathematical languages.
However, both have proven to be very useful to tackle quantum problems,
providing us with different perspectives and/or strategies on the same
issue.

In general, once a quantum picture is accepted, even if it makes the same
predictions as other available pictures, there are at least two intertwined
aspects that set a difference:
\begin{itemize}
 \item[(1)] A {\it physical aspect}. This is analogous to what happens
 in classical mechanics, where each (classical) picture allows us to
 approach the same physical problem from a different perspective,
 highlighting a particular feature or concept, and thus providing us
 with alternative insights -- think, for instance, of Newton's forces
 at a distance versus Faraday's fields.
 This aspect, connected to the way how we explain and understand
 physical phenomena, has an intrinsic interest (interpretational) at
 the fundamental level.

 \item[(2)] A {\it practical aspect}. It is associated with the potentiality
 of the picture as a resource to generate and develop new numerical
 tools based on it, aimed at solving in a more efficient way quantum
 problems.
 This important feature allows us to recreate (simulate) quantum
 phenomena and, therefore, to extend the theory beyond the limitations
 of analytical treatments, which is the case in most cases of
 practical interest.
 Usually this is precisely the aspect behind the fact that a given
 picture will eventually become or not widely accepted (or, at least,
 it has an important weight on this fact).
\end{itemize}
It is by virtue of these two aspects that we do not need to chose
between one approach and the rest.
Same happens with Bohmian mechanics.
In this particular case, apart from its intrinsic physical interest,
the practical side is also backed by the number and variety of
numerical algorithms appeared since the end of the
1990s \cite{wyatt:prl:1999}, the so-called {\it quantum trajectory
methods} \cite{wyatt-bk}.
These methods take advantage of an earlier reformulation of
Schr\"odinger's equation in a hydrodynamic form, proposed in 1926 by
Madelung \cite{madelung:ZPhys:1926}, which establishes a link between
quantum mechanics and classical hydrodynamics, and hence an effective
transfer of numerical resources from the latter to the former.

The second criticism is related to the above physical aspect,
specifically the fact that Bohmian mechanics is based on the concept
of trajectory, incompatible in principle with quantum mechanics --
Feynman's path integral approach also found an analogous
opposition from Bohr (see Ref.~\cite{Mehra-Feynman-bk},
pp.~245--248).
The apparent ``harm'' here comes from a direct association of Bohm's
trajectories, understood as ``hidden variables'' (not experimentally
accessible), with ``real'' paths followed by the quantum
system \cite{note3} (quantum particle).
This association arises from the first of the two papers that Bohm
published \cite{bohm:PR:1952-1} in 1952, where he states:
\begin{quote}
 The usual interpretation of the quantum theory is self-consistent,
 but it involves an assumption that cannot be tested experimentally,
 {\it viz.}, that the most complete possible specification of an
 individual system is in terms of a wave function that determines only
 probable results of actual measurement processes.
 The only way of investigating the truth of this assumption is by trying
 to find some other interpretation of the quantum theory in terms of at
 present ``hidden'' variables, which in principle determine the precise
 behavior of an individual system, but which are in practice averaged
 over in measurements of the types that can now be carried out.
 In this paper [\ldots] an interpretation of the quantum theory in
 terms of just such ``hidden'' variables is suggested.
\end{quote}
This identification has been used to raise Bohmian mechanics to the
level of ontological quantum theory, where appealing to the action of
an external observer to cause the collapse of the wave function is not
necessary -- a quantum theory without
observers \cite{goldstein:phystoday:1,goldstein:phystoday:2}
However, apart from just being a matter of interpretation, there is no
formal or empirical evidence for such a connection.
Bohmian trajectories only serve to monitor the flux or diffusion of the
quantum system throughout the corresponding configuration
space \cite{sanz:AJP:2012}.


\subsection{Quantum mechanics and hidden variables}
\label{sec22}

\begin{figure}[t]
\begin{center}
 \includegraphics[width=8.25cm]{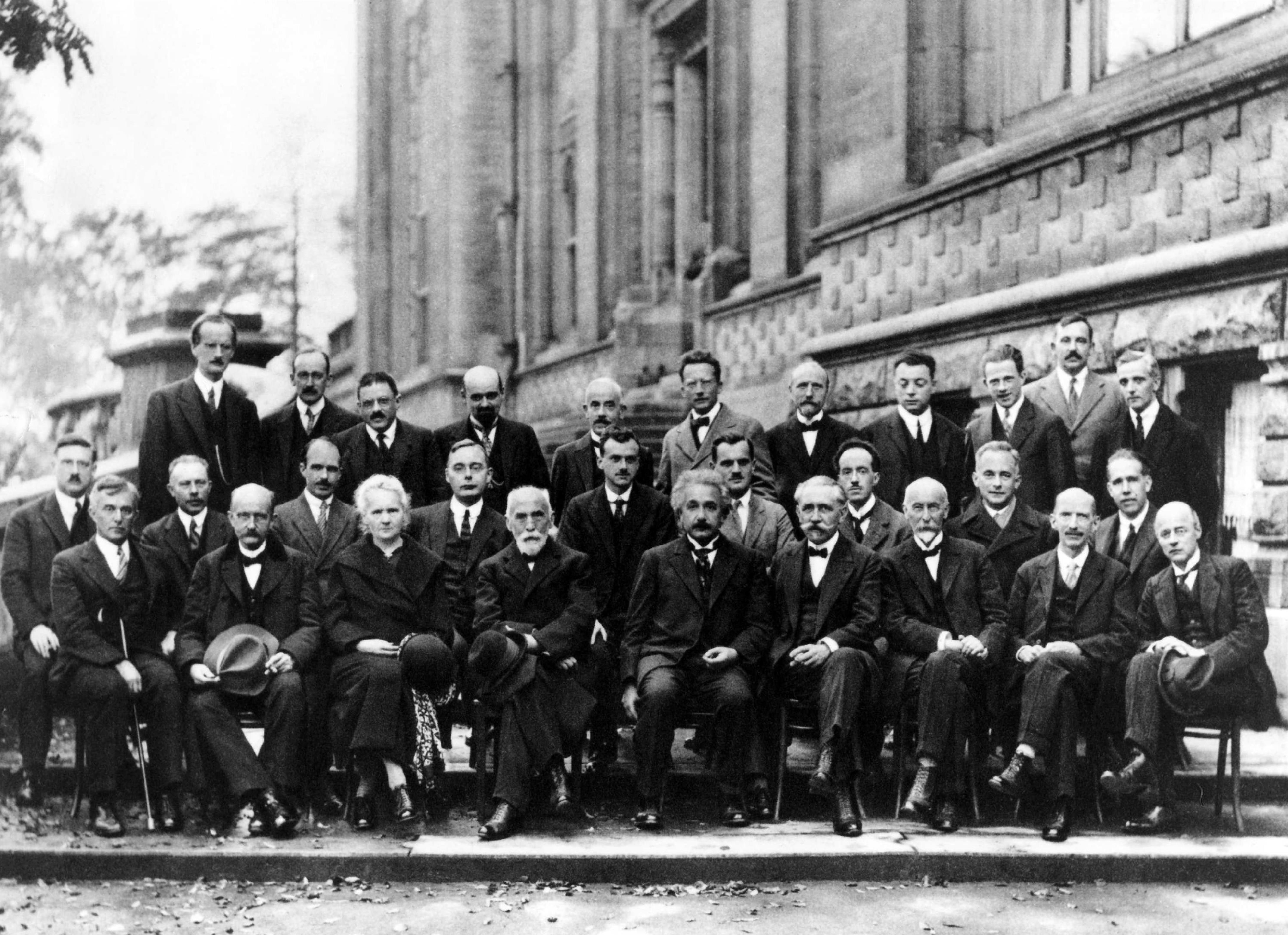}
 \caption{Attendees to the 5th International Solvay Conference
 (Brussels, 1927).}
 \label{fig1}
\end{center}
\end{figure}

The original contextual background of Bohm's approach is the debate
on the interpretation of the wave function \cite{jammer-bk:1966,zurek-bk,hawking-bk},
specifically the quest for a realistic description of quantum
phenomena -- an {\it ontological interpretation} of the wave function
-- led by Einstein in the 1930s.
It is thus worth getting back to the 5th International Solvay
Conference \cite{marage-wallenborn-bk}, ``Electrons et photons'', held
in Brussels in 1927, probably the most famous one within this
conference series.
The attendees to this conference (see Fig.~\ref{fig1}) addressed the
problem of the interpretation of quantum mechanics, an issue that was
somehow settled by Bohr (with no little opposition from Einstein ---
``God does not play dice'') and eventually gave rise to the so-called
Copenhagen or orthodox interpretation \cite{cushing-bk:1994}.
According to this interpretation, it does not make any sense to ask
what or where a quantum system is until a measurement is performed.
This way to conceive the quantum realm is tightly associated with the
rather abstract matrix formulation of quantum mechanics, formulated
by Heisenberg, Born, and Jordan in 1925 \cite{waerden-bk}.
No need to say that this position did not favor at all the hypothesis
presented in the same conference by de Broglie on the possibility of a
pilot wave guiding the quantum system, strongly contested by Pauli and
other participants \cite{bacciagaluppi-valentini-bk}.

Copenhagen left little room for a different view than the one posed
by Bohr: unlike classical mechanics, quantum mechanics only provides us
with probabilistic information on the system, formally accessible
through the squared modulus of the wave function describing the actual
state of such a system.
This was opposite to Einstein's view: the wave
function, by its own, cannot completely specify the state of the system,
hence a set of additional parameters or variables are needed
to unambiguously specify the state of the system, and therefore to
provide of a full physical sense to the statistical outcomes rendered
by the wave function.
In brief, the wave function description is incomplete.
These variables are referred to as {\it hidden variables}, because
in principle they would be ``hidden'' to the experimentalist (not
experimentally accessible).

At a formal level, von Neumann showed (see p.~210 in
Ref.~\cite{vonNeumann-bk:1932En}) that ``an introduction of
hidden parameters is certainly not possible without a basic change in
the present theory'' (quantum mechanics).
To illustrate the role played by the hidden variables, von Neumann
recalls the probabilistic way how the kinetic theory of gases operates,
where the set of Newtonian positions and momenta that uniquely specify
the state of each gas atom or molecule is replaced by only two
statistical parameters: pressure and temperature.
However, unlike this classical picture, von Neumann proves (see
pp.~323-325 in Ref.~\cite{vonNeumann-bk:1932En}) that quantum
mechanics cannot be re-derived as a statistical approximation of an
underlying causal classical-like theory based on a set of additional
variables.
It is this incompatibility what leads to the impossibility to reach
a more complete specification of the state of a quantum system than
the one provided by the associated wave function.

Twenty years after von Neumann's proof on the impossibility of hidden
variables, Bohm found \cite{bohm:PR:1952-1,bohm:PR:1952-2} a
counterexample after recasting Schr\"odinger's equation in the form
of two real-valued partial differential equations (see Sec.~\ref{sec3}).
This model was formally identical to the hydrodynamic one proposed in
1926 by Madelung \cite{madelung:ZPhys:1926}, although interpretively
in the spirit of de Broglie's pilot wave theory \cite{broglie-bk:1960}.
According to Bohm, a causal explanation to the quantum outcomes
observed is possible in terms of statistical distributions of initial
conditions in configuration space for quantum systems.
He associated these initial conditions with the non-observable hidden
variables referred to by von Neumann, thus proving a way to recast
quantum mechanics in terms of a set of hidden variables compatible
with all its predictions.
Causality and probability could then be unified within quantum
mechanics without the formulation of a totally new theory, as claimed
by von Neumann.

Bohm's proposal could reproduce all the predictions of quantum
mechanics, although it also introduced a disturbing element into play:
{\it nonlocal} hidden variables.
Classical systems are ruled by the {\it principle of locality}, i.e., no
influence on a given physical system can travel faster than the speed of
light.
Therefore, any action on these systems must come from a certain neighborhood.
In quantum mechanics, though, things work differently and actions at remote
distances may cause important disturbances on the system without
violating the special theory of relativity, as it was pointed out in
1935 by Einstein, Podolsky and Rosen \cite{EPR:PhysRev:1935} as well as
by Schr\"odinger \cite{schrodinger:ProcCamPS:1935,schrodinger:ProcCamPS:1936}.
This phenomenon was possible by virtue of the property of
{\it entanglement}.
Accordingly, as noticed by Schr\"odinger, as soon as two quantum
systems interact, they will remain strongly correlated independently
of the distance between them \cite{schrodinger:ProcCamPS:1935}; any
measurement performed on one of the systems will instantaneously
determine the outcome of the same measurement on the other one.
This puzzling behavior, with no classical counterpart, started a
longstanding debate on the incompleteness of the wave function and the
necessity to explain such a ``spooky action at distance'' in terms of
hidden variables.
Within this scenario, Bohm's work did not satisfy either those against
hidden variables or those in favor.
His contribution was considered to be wrong and, after a few years,
it was essentially relegated to oblivion.

In the 1960s Bell got back to those fundamental
questions \cite{bell:physics:1964,bell:RMP:1966} (precisely after
coming across Bohm's papers \cite{gilder-bk}), reformulating the
problem of hidden variables and changing the landscape of quantum
mechanics by laying the grounds for what is now known as the Second
Quantum Revolution \cite{milburn:PTRSLA:2003}.
Specifically, what Bell found is that no physical theory of local
hidden variables can ever reproduce all the predictions of quantum
mechanics (see p.~542 in Ref.~\cite{parker-bk:1994}), because
quantum mechanics is {\it intrinsically nonlocal}.
Bohm was not wrong indeed; in spite of Einstein's reluctancy
against the idea of quantum mechanics being nonlocal, Bohm provided a
neat model that explained this quantum feature with the presence of
a nonlocal potential, the so-called quantum potential (see
Sec.~\ref{sec3}).
This potential is implicit in Schr\"odinger's equation and becomes
explicit when it is recast in a Hamilton-Jacobi form by means of a
nonlinear transformation from complex to real field variables.
Because von Neumann's theorem implicitly assumes non-contextuality (von
Neumann had in mind ensembles in classical-like variables), it can only be
correctly applied to {\it local} hidden-variable models, leaving aside a
distinctive aspect of quantum mechanics, namely its nonlocality, which is
made more apparent through Bohm's formulation.
The story that follows Bell's discovery, particularly after its
experimental confirmation in 1981 and 1982 by Alain Aspect and
co-workers \cite{aspect:PRL:1981,aspect:PRL:1982-1,aspect:PRL:1982-2}
is well known \cite{note4}: the appearance and fast development of a series
of new areas of quantum mechanics with very important technological
applications, such as quantum teleportation, quantum computing, quantum
cryptography, quantum imaging or quantum sensors.

Apart from inspiring Bell, Bohm himself went back in the 1970s to his
former works from 1952 in collaboration with his colleague Hiley, who
had already produced with PhD students Chris Dewdney \cite{note1new}
and Chris Philippidis two seminal works based on
Bohm's approach, one about Young's two-slit experiment \cite{dewdney:NuovoCimB:1979}
and another one about scattering of square potential wells and barriers
\cite{hiley:foundphys:1982}.
They started working on the implementation of the physics behind Bohm's
suggested interpretation \cite{hiley:SHPMP:1997}, stressing the role of
the quantum potential as an interpretational tool to explain well-known
quantum phenomena in a causal fashion.
A number of papers followed along the 1980s and early 1990s by the
members of this group as well as by other authors, where trajectories
were produced for different quantum problems, proving the correctness
of Bohm's ideas.
Much of that work is summarized in the monograph {\it The Quantum
Theory of Motion} \cite{holland-bk}, published in 1993 by Peter
Holland; other more fundamental interpretational aspects are discussed
in {\it The Undivided Universe} \cite{bohm-hiley-bk}, published in
1993 by Bohm and Hiley.

By the beginning of the 1990s, another Bohmian school arose around
Detlef D\"urr, Sheldon Goldstein and Nino Zhang\`{\i}, where Bohmian
mechanics was approached through a statistical mechanical perspective
instead of the usual quantum potential one, introducing the notion of
the quantum equilibrium hypothesis \cite{duerr:JStatPhys:1992a}.
According to this idea, the quantum states we observe are states
at equilibrium, i.e., states that result from the action of a sort of
subquantum medium on arbitrary states that evolve or ``thermalize''
into the usual wave functions that satisfy Schr\"odinger's equation.
It is at this point where Bohmian mechanics enters as a natural way to
explain how such a thermalization process would take place, in analogy
to how the thermalization of classical ensembles produces the
statistical Boltzmann distributions we observe at equilibrium.

\begin{figure}[t]
\begin{center}
 \includegraphics[width=8.25cm]{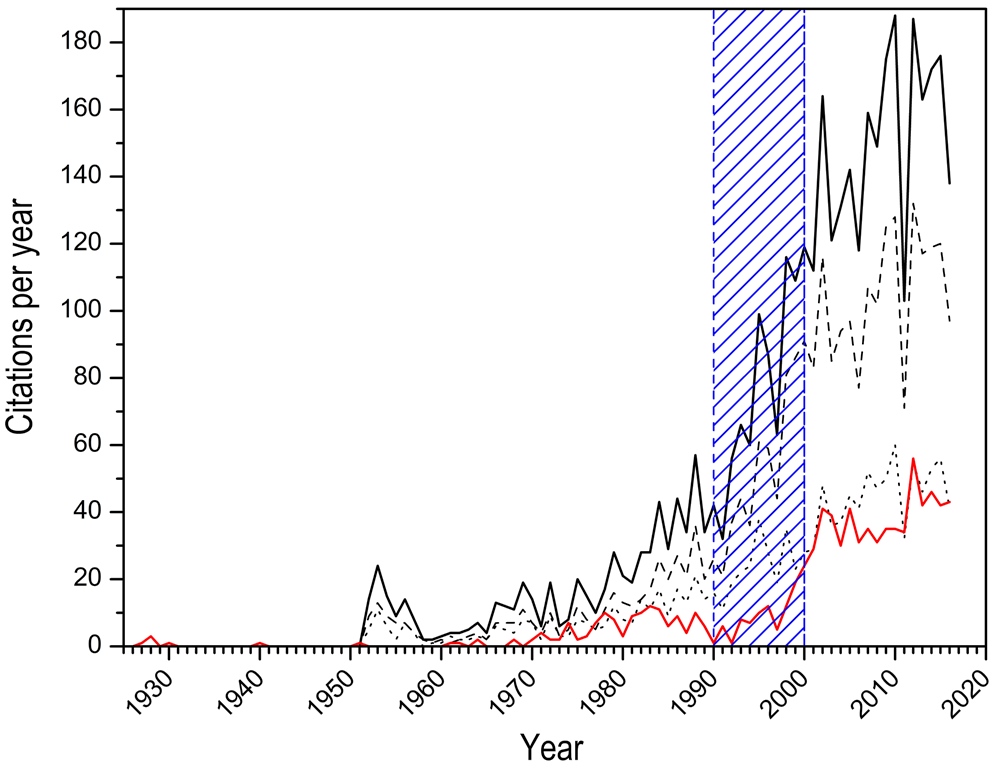}
 \caption{Number of citations per year of the two papers published by
  Bohm in 1952 (black line) and the paper published by Madelung in 1926
  (red line).
  The black dashed and dotted lines make reference, respectively, to
  the first and second Bohm papers separately.
  The vertical blue band serves to denote a kind of transition period
  where the role of Bohmian mechanics as a working tool (both
  computational and interpretive) started gaining importance to the
  detriment of other more fundamental (hidden-variable related)
  aspects.}
 \label{fig2}
\end{center}
\end{figure}

From those early days to date the use of Bohmian mechanics has
remarkably increased to tackle different quantum mechanical
problems \cite{wyatt-bk,duerr-bk:2009,duerr-bk:2013,chattaraj-bk,%
hughes-bk,sanz-bk-2,oriols-bk,benseny:EPJD:2014}.
Figure~\ref{fig2} may help to get an idea on the acceptance of
Bohmian mechanics.
In this figure we observe the time-evolution of the amount of works
citing Bohm's seminal works from 1952 with black line: dashed line
for paper I, dotted line for paper II, and solid line for the total
number of citations (papers I \& II).
Data have been taken from the ISI Web of Science.
Although there might be some other works dealing with or based on Bohmian
mechanics that do not cite these two papers, they still constitute a
reliable reference of the trend over time, which is what matters at
this point.
The case of de Broglie's pilot wave model has not been considered here
because, given that there are several possible references (published in
different periods), the follow-up is more complicated and probably not
that much reliable.
As can be seen, since the 1990s the interest in Bohmian mechanics
has remarkably increased up to date, when it is more broadly considered
and accepted among its users as a computational and interpretive
working tool, to tackle a wide variety of problems in different areas
of physics \cite{wyatt-bk,chattaraj-bk,hughes-bk,sanz-bk-2,oriols-bk,benseny:EPJD:2014},
than as a hidden-variable theory, i.e., an ontological interpretational
model to understand the quantum realm, as it was essentially the case
from the 1950s to the 1980s.

Due to its interest in the development of the ideas implicit
in Bohmian mechanics (the use of streamlines as a computational and
interpretive resource), the evolution of the citations to the paper
published by Madelung in 1926 is also displayed in Fig.~\ref{fig2}
with red solid line.
Nonetheless, it should be noticed that there might be many more works
based on quantum hydrodynamics that those citing Madelung's paper,
since it is an important approach in areas such as plasma physics, particle
physics, condensed matter physics, superfluidity, etc., used and promoted,
among others, by Landau or London concerning the modeling of liquid
helium in the 1940s \cite{landau:JPUSSR:1941,london:RMP:1945,ziman:PRSL:1953,bierter:JLTP:1969}.
In these latter cases, neither there is a direct link to Bohmian mechanics, and
still the essence of such formulations are exactly the same.

Although Madelung's hydrodynamic formulation was neglected until the
1970s, from this time onwards it started being considered as a practical tool
to visualize the evolution in real time of intermediate stages of processes
with interest in Chemistry, from molecular
reactions \cite{mccullough:JCP:1969,mccullough:JCP:1971-1,mccullough:JCP:1971-2}
to the development and evolution of vortex
dynamics \cite{hirsch:JCP:1974-1,hirsch:JCP:1974-2,hirsch:JCP:1976-1,hirsch:JCP:1976-2,hirsch:JCP:1977}
(with the concept of quantum vortex dating back to the works on superfluid helium
from the 1940s and 1950s).
The idea behind these simulations was to provide a full and deeper
understanding of the processes investigated without the need to use
interpretations based on classical trajectory computations (quantum-classical
correspondence).
This idea run through the 1980s and 1990s, and was further extended to
the analysis of magnetism in molecular systems
\cite{lazzeretti:CPL:1981,lazzeretti:IJQC:1984,lazzeretti:PNMRS:2000,lazzeretti:IJQC:2011,gomes:JCP:1983-1,gomes:JCP:1983-2,berger:AngewChem:2010}.
However, the most noticeable boost arises by the end of the 1990s, when
Courtney Lopreore and Robert ``Bob'' Wyatt proposed \cite{wyatt:prl:1999} the first {\it quantum
trajectory method} \cite{wyatt-bk}, and Madelung's approach merges with
Bohmian mechanics (except, perhaps, in the area of the quantum
foundations, where there are still some objections to such a
unification), explaining the rapidly increasing number of citations
from the 1990s onwards in Fig.~\ref{fig2}.

This section would be incomplete without a mention to the experiments
with classical vibrating fluids performed by Yves Couder and Immanuel
Fort \cite{couder:Nature:2005,couder:PRL:2006,couder:JFluidMech:2006,couder:PNAS:2010}
at the Universit\'e Paris Diderot, and John Bush \cite{bush:PNAS:2010,bush:PRE:2013,bush:ARFM:2015}
at the MIT, which show a nice classical counterpart of the behavior
devised by de Broglie with respect to quantum systems through his
pilot-wave theory: a fluid droplet bounces and generates a
self-sustained wave that, at the same time, guides the motion of the
droplet \cite{note5}.
Based on these experiments and de Broglie's former double-solution
program, Thomas Durt has recently proposed \cite{durt:EPL:2016} a model
to explain them that assumes the existence of a nonlinear self-focusing
gravitational-like potential that prevents the particle from undergoing
spreading (it is kept as a peaked soliton).


\section{Quantum mechanics within the Bohmian picture}
\label{sec3}


\subsection{Basic formal aspects}
\label{sec31}

The basic equations of Bohmian mechanics can be obtained in different
ways.
The usual one departs from recasting the wave function in polar
form \cite{madelung:ZPhys:1926,bohm:PR:1952-1,holland-bk},
\be
 \Psi ({\bf r},t) = \sqrt{\rho ({\bf r},t)}\ \!
   e^{iS({\bf r},t)/\hbar} .
 \label{eq2}
\ee
This is just a nonlinear transformation from the complex field
variables $\Psi$ and $\Psi^*$ to the real field variables $\rho$ and
$S$ (from now on the explicit dependence on ${\bf r}$ and $t$ will be
omitted for simplicity).
In particular,
\be
 \rho = \Psi \Psi^*
 \label{eq3}
\ee
is the usual probability density, and
\be
 S = \frac{\hbar}{2i}\ \! \ln \left( \frac{\Psi}{\Psi^*} \right)
 \label{eq4}
\ee
provides the local value of the system's quantum phase.
Physically, the probability density represents the statistical
distribution of possible realizations (detections) of a given quantum
system on a certain region $d{\bf r}$ of the configuration space
at a given time $t$, e.g., a statistically meaningful number of counts
registered by a detector positioned at a certain angle from a target
for a given exposure time, for instance.
Because it can be measured (intensity) and has an associated operator
(the density operator), we call it a {\it quantum observable}.
Regarding the quantum phase $S$, it provides us with information
about the local variations of the system quantum phase, which is not a
quantum observable, because it can only be inferred through indirect
measurements \cite{note6}, e.g., an interference pattern.

After substitution of the polar ansatz (\ref{eq2}) into the time-dependent
Schr\"odinger equation,
\be
 i\hbar \ \! \frac{\partial \Psi}{\partial t}
  = -\frac{\hbar^2}{2m}\ \! \nabla^2 \Psi + V \Psi ,
 \label{schro}
\ee
we obtain the equations of motion that describe the time-evolution of
the real-valued field variables,
\ba
 \frac{\partial \rho}{\partial t} & + &
  \nabla \left(\rho\ \! \frac{\nabla S}{m}\right) = 0 ,
  \label{eq5} \\
 \frac{\partial S}{\partial t} & + &
  \frac{(\nabla S)^2}{2m} + V + Q = 0 ,
 \label{eq6}
\ea
with
\ba
 Q & = & - \frac{\hbar^2}{8m} \left[
  2 \left( \frac{\nabla^2 \rho}{\rho} \right)
  - \left( \frac{\nabla \rho}{\rho} \right)^2 \right]
 \nonumber \\
 & = & - \frac{\hbar^2}{2m} \left\{
  {\rm Re} \left( \frac{\nabla^2 \Psi}{\Psi} \right)
  + \left[ {\rm Im} \left( \frac{\nabla \Psi}{\Psi} \right)
     \right]^2 \right\}
 \label{eq7}
\ea
being the so-called {\it quantum potential}.
Equation (\ref{eq5}) is readily identified with the usual continuity
equation; Eq.~(\ref{eq6}) is regarded as a quantum Hamilton-Jacobi
equation due to its close resemblance with its classical analog.
Notice that the quantum phase $S$ has the same dimensions as the
classical mechanical action, [energy]$\times$[time], which is a
reminiscence of the Hamiltonian analogy considered by Schr\"odinger
to derive his equation.
Regarding the quantum potential $Q$, it has typically assigned the
role of a potential function; $Q$ and $V$ thus form a global effective
potential acting on the trajectories.
However, $Q$ actually comes from the action of the
Laplacian operator, $\nabla^2$, on the wave function in Eq.~(\ref{schro})
and therefore it is closer to a kinetic-like energy than to a potential
function.
This question will be revisited again in Sec.~\ref{sec4}.

Once a Hamilton-Jacobi equation is introduced, as it is done in
classical mechanics, the concept of trajectory arises in a natural way
if we take into account that such trajectories are characteristic
solutions obtained after integrating the quantum version of the Jacobi
law,
\be
 \dot{\bf r} = \frac{\nabla S}{m} ,
 \label{eq8}
\ee
with initial condition ${\bf r}_0$.
This simple idea is precisely what Bohm suggested in 1952 and, at the
same time, the way how the hidden-variable controversy comes into
play, since after this postulate one feels very much tempted to think
(and identify) these trajectories with the actual paths followed by a
quantum particle.
Notice that this is not in contradiction with Heisenberg's uncertainty
relations, because uncertainty comes from the fact that,
in practice, it is impossible to accurately determine $x_0$; initial
conditions are randomly distributed, with the ensemble statistics being
described by $\rho$ at $t=0$.

In general, the Bohmian set of equations has been approached or
considered in the literature in two different ways, each one with a
different purpose.
According to Wyatt \cite{wyatt-bk}, these two approaches can be denoted
as {\it analytic} and {\it synthetic schemes}, which closely resemble
the homologous analytic-synthetic distinction or dichotomy in
Philosophy to primarily classify (logical) propositions or judgments,
one of the pillars of Kant's philosophical system \cite{kant:1781}.
In the first case, first the Schr\"odinger equation is solved and, once we
get the wave function, the trajectories are computed and subsequently
used to analyzed, interpret, understand and explain the system under study.
In the second case, the calculation of the trajectories is independent
of the wave function, thus being the optimal route considered in the
design of new quantum propagation methods based on Bohmian
mechanics.

In order to reach Eq.~(\ref{eq8}) and hence to associate or
compute trajectories that monitor the evolution of a quantum system,
it is not necessary introducing an additional postulate in quantum
mechanics, since in its conventional Schr\"odinger formulation we
already have all the necessary elements.
Specifically, we have the probability density and also a current
density \cite{schiff-bk},
\be
 {\bf J} = \frac{1}{m}\ \!
  {\rm Re} \left\{ \Psi^* \hat{\bf p}\ \! \Psi \right\} ,
 \label{eq9}
\ee
with $\hat{\bf p} = - i\hbar\nabla$ being the usual momentum operator.
This quantity allows us to introduce in a natural way a velocity field
variable, ${\bf v} = {\bf J}/\rho$, which formally coincides with the
Bohmian velocity given above,
\be
 {\bf v} = \frac{\bf J}{\rho} = \frac{1}{m}\ \!
  {\rm Re} \left\{ \frac{\hat{\bf p}\ \! \Psi}{\Psi} \right\}
  = \frac{\nabla S}{m} = \dot{\bf r} ,
 \label{eq10}
\ee
although without appealing to a hidden-variable scenario.
The fact that all quantities to the right of the first and
second equalities are well-defined within the conventional version of
quantum mechanics, and hence the use of the quantities to the right of
the third and fourth equalities, namely $S$ and ${\bf r}$, should also
be legitimately considered, leaving aside further interpretational
issues.
Actually, taking into account the definition of $\hat{\bf p}$, the
quantum potential can be recast as
\ba
 Q & = & \frac{1}{m} \left\{ \frac{1}{2}\ \!
  {\rm Re} \left( \frac{\hat{\bf p}^2 \Psi}{\Psi} \right)
  + \frac{1}{2}\ \! \left[
    {\rm Im} \left( \frac{\hat{\bf p} \Psi}{\Psi} \right)
     \right]^2 \right\} ,
 \label{eq7b}
\ea
which further stresses the dynamical origin of $Q$ mentioned above.

To conclude this section, it is worth stressing that analogous approaches
to what we now know as Bohmian mechanics have been used in the
literature to solve different problems.
For example, in Optics we find models based on the Poynting vector and
the Maxwell equations already in the 1950s \cite{braunbek:Optik:1952,bornwolf-bk},
and only much later the approaches of Bohm and Madelung were
considered \cite{prosser:ijtp:1976-1,prosser:ijtp:1976-2,dahmen:AnnPhys:1998,zakowicz:PRE:2001,herrmann:AJP:2002,hesse:JQSRT:2008,sanz:AnnPhysPhoton:2010,bliokh:NewJPhys:2013}.
In Acoustics the use of streamlines was also suggested in the 1980s as
a visualization and analysis tool \cite{waterhouse:JASA:1985,waterhouse:JASA:1986-1,waterhouse:JASA:1986-2,waterhouse:JASA:1987-1,waterhouse:JASA:1987-2},
although there was not a direct link with Madelung or Bohm in this
direction until recently \cite{milena:JRLR:2015}.
Finally, we also find different models aimed at treating dissipation in
quantum systems, such as the one proposed by Kostin \cite{kostin:jcp:1972},
which is based on the idea of adding nonlinear
contributions to the Schr\"odinger equation, which, interestingly, are
given in terms of the phase of the wave function, and therefore the
Bohmian momentum (which is identified with the dissipative term that
appears in the classical Newton equations of motion with friction).


\subsection{A simple illustration: interference from two mutually
coherent sources}
\label{sec4}

Quantum interference is one of the distinctive traits of quantum mechanics,
providing a rather convenient environment to investigate fundamental questions
by devising different types of interferometric experimental setups
\cite{yuan:LSA:2016,zhou:SciBull:2017,long:SciChina:2018,li:EPL:2017}.
Therefore, it is not strange that interference has been a recurrent topic in the
Bohmian literature since the former numerical simulation published by Philippidis,
Dewdney and Hiley in the case of Young's two-slit experiment \cite{dewdney:NuovoCimB:1979}.
Thus, for a proper understanding of the role of Bohmian mechanics within the
usual context of quantum mechanics, we are now going to consider a
simple working model on Young's two-slit experiment.
This model combines in a simple way analyticity and numerics with the
purpose to be easily considered in the classroom.
To start with, we consider paraxial conditions \cite{sanz:AOP:2015}, so
that interference is assumed to happen essentially along the transverse
direction (i.e., parallel to the screen where the two slits are
supposed to be), here denoted by the $x$ coordinate.
Accordingly, only this (transverse) component of the wave function is
considered.
Furthermore, we want to focus on the phenomenology of the experiment,
namely the appearance of interference by coalescence of two waves
coming from {\it two mutually coherent sources}, so the time-evolution
will start right behind the two slits.
To avoid edge-related diffraction effects, the initial state is
described by a coherent superposition of two Gaussian wave packets,
\be
 \Psi(x,0) \sim e^{-(x - d/2)^2/4\sigma_0^2}
              + e^{-(x + d/2)^2/4\sigma_0^2} ,
 \label{eq0}
\ee
which in the far field provides us with a neat interference pattern
only affected by a Gaussian envelope (compare this with the sinc
function affecting square waves \cite{sanz:JPCM:2002}).
The time-evolution of (\ref{eq0}) is fully analytical \cite{sanz-bk-2};
at a time $t$ it is easy to show that the wave function reads as
\be
 \Psi(x,t) \sim e^{-(x + d/2)^2/4\sigma_0\tilde{\sigma}_t}
              + e^{-(x - d/2)^2/4\sigma_0\tilde{\sigma}_t} ,
 \label{eq1}
\ee
with
\be
 \tilde{\sigma}_t = \sigma_0
  + i\ \! \left(\frac{\hbar}{2m\sigma_0}\right) t .
 \label{complex}
\ee
Notice in both (\ref{eq0}) and (\ref{eq1}) that the global phase and
normalizing factors have been neglected.
This is only to simplify the analysis, since they are irrelevant from
a quantum dynamical viewpoint, as can be seen below.

The appearance of interference can be readily understood with this
simple model by just noticing in (\ref{eq1}) the appearance of the
time-developing complex factor (\ref{complex}) in the argument of both
exponentials.
Taking this factor into account, these exponential functions can be
recast as
\be
 e^{-(x \pm d/2)^2/4\sigma_0\tilde{\sigma}_t} =
  e^{-(x \pm d/2)^2/4\sigma_t^2}
  e^{i(\hbar t/8m\sigma_0^2)(x \pm d/2)^2/4\sigma_t^2} ,
 \label{dec}
\ee
with
\be
 \sigma_t \equiv |\tilde{\sigma}_t| =
  \sigma_0 \sqrt{1 + \left(\frac{\hbar t}{2m\sigma_0^2}\right)^2} ,
 \label{sigma}
\ee
The first exponential function on the right-hand side of Eq.~(\ref{dec})
describes the widening undergone by the wave packets as time
proceeds; the second one accounts for the development
of a phase factor depending on time and, more importantly, the $x$
coordinate.
This factor induces the appearance of a phase field, which
assigns a value to the ({\it local}) phase at each space point and
time.
Eventually, this leads to the appearance of interference features, even
in the very case that there is no any relative phase between the two
wave packets at $t=0$.
This is more apparent through the associated probability density,
\ba
 \rho(x,t) & \sim & e^{-(x + d/2)^2/2\sigma_t^2}
  + e^{-(x - d/2)^2/2\sigma_t^2}
  \nonumber \\
  & & + 2 e^{-[x^2 + (d/2)^2]/2\sigma_t^2}
  \cos \left[ \left( \frac{\hbar t d}{4m\sigma_0^2\sigma_t^2}\right)
   x \right] .
 \nonumber \\ & &
 \label{probdens1}
\ea
%
%
This result shows that the total intensity recorded
is not the bare addition of partial intensities, as in classical
statistical mechanics.
Nonetheless, more interestingly, this model also shows that
interference features are present since the very beginning through
the third oscillatory factor, even if the initial distance between
both wave packets is too large compared to their widths (i.e.,
$d \gg \sigma_0$).
The appearance of this term as soon as $t \neq 0$, no matter
how relevant it is, is what allows us to talk about and formally
describe as {\it coherence} when dealing with the superposition
of waves.

\begin{figure*}[!ht]
\begin{center}
 \includegraphics[width=15cm]{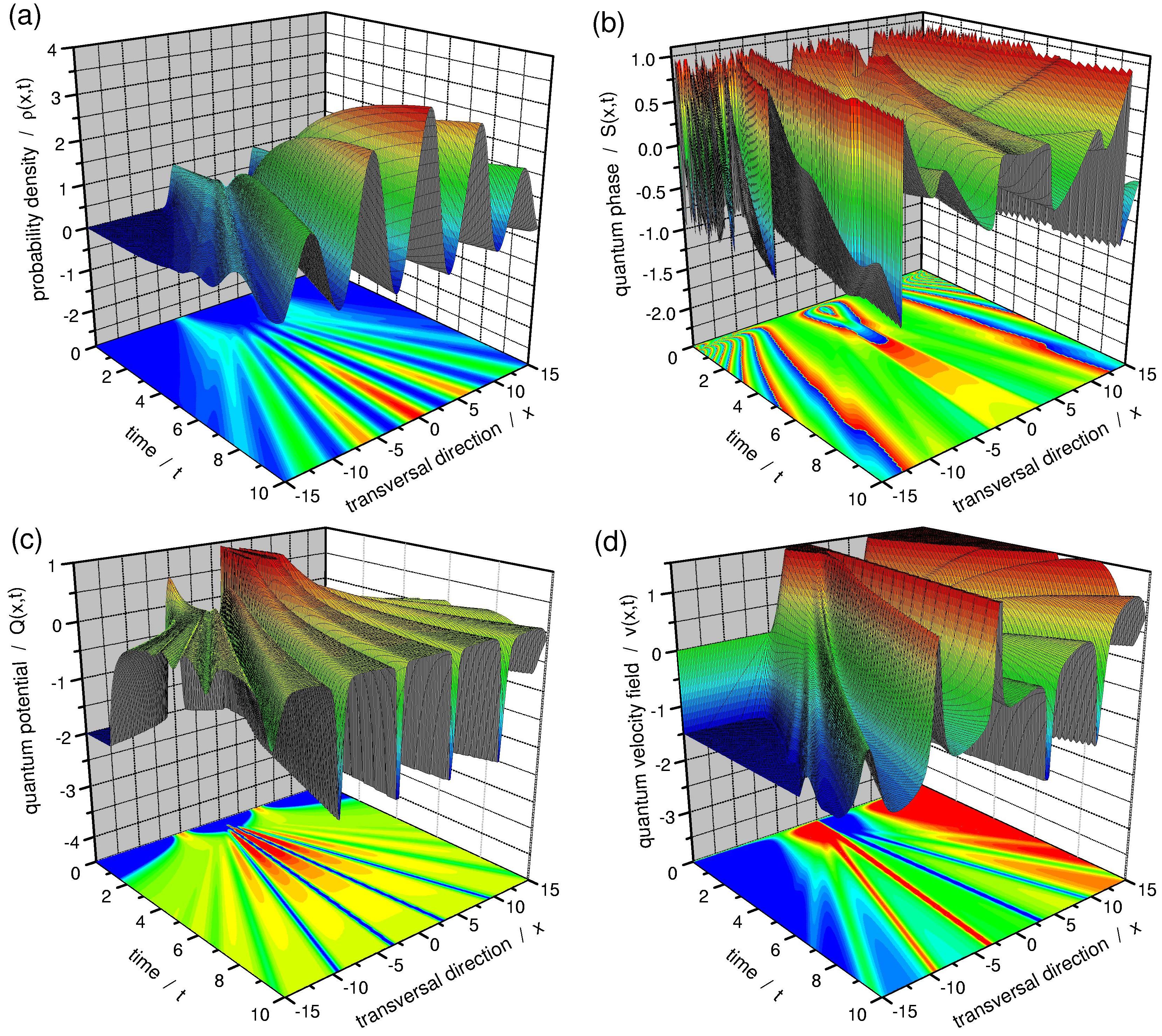}
 \caption{Numerical simulation illustrating different aspects of
  Young's two-slit experiment \cite{sanz:foundphys:2015}.
  The contour-plots illustrate the evolution in time of the probability
  density (a), the quantum phase (b), Bohm's quantum potential (c), and
  the quantum velocity field determined by the wave function (d).
  The black solid lines in all plots represent ensembles of Bohmian
  trajectories leaving the slits with different initial conditions
  and, accordingly, ending at different points of a scanning screen
  (detector).}
 \label{fig3}
\end{center}
\end{figure*}

The time-evolution of the probability density (\ref{probdens1}) is
displayed in Fig.~\ref{fig3}(a) in terms of surface and contour plots.
The numerical values in this simulation are $\hbar =1$, $m=1$,
$\sigma_0 =0.5$, and $d=10$, and the evolution been halted at $t=10$,
enough to observe all the dynamical regimes ruling the behavior of the
quantum system \cite{sanz:AJP:2012}.
These particular values have been chosen, because they produce a
characteristic time ruling the wave-packet spreading dynamics of
$\tau = 2m\sigma_0^2/\hbar = 0.5$ (at $t=\tau$,
$\sigma_t = \sqrt{2}\sigma_0$), which means that at the time when
interference fringes start appearing both wave packets already display
a linear widening \cite{sanz:AJP:2012}.
As it can be noticed, until $t \approx 2$, corresponding to the
Huygens-Ehrenfest and Fresnel regimes that characterize the early
stages of a wave function evolution, the wave packets seem to evolve
independently of each other.
Then, for $2 < t < 4$, a fringed structure starts developing until
$t \approx 5$, when such a structure remains stationary, with nodes
that spread out both sides linearly with time.
This latter stage corresponds to the Fraunhofer regime, where we can
neatly observe the characteristic fringed pattern of Young's
experiment.
This long-time limit can be easily described with our working model.
Notice that asymptotically, for $t \gg \tau$, Eq.~(\ref{sigma}) can be
approximated by $\sigma_t \approx \hbar t/2m\sigma_0$.
The probability density (\ref{probdens1}) then reads as
\ba
 \rho(x,t) & \sim & e^{-(x + d/2)^2/2(v_s t)^2}
  + e^{-(x - d/2)^2/2(v_s t)^2}
  \nonumber \\
  & & + 2 e^{-[x^2 + (d/2)^2]/2(v_s t)^2}
  \cos \left[ \left( \frac{m d}{\hbar}\right) \frac{x}{t} \right] ,
 \nonumber \\ & &
 \label{probdens2}
\ea
%
%
which is modulated by an even simpler phase factor, and where
$v_s \equiv \hbar/2m\sigma_0$ is the spreading rate or velocity at
which Gaussian wave packets widen asymptotically \cite{sanz:JPA:2008}.
According to the phase factor in (\ref{probdens2}), the distance
between adjacent nodes or maxima increases nearly linearly with time
as
\be
 \Delta x = \left( \frac{2\pi\hbar}{md} \right) t .
 \label{nodes}
\ee
In our case, this linearity condition is satisfied precisely for
$t \approx 5$ ($\gg \tau$).
In particular, for $t=10$ we have $\Delta x = 2\pi$, in
compliance with the results displayed in Fig.~\ref{fig3}(a).

The probability density conveys information on the chance to
find the particle within a certain region of the configuration space
after it (its associated wave function) has passed through the slits.
At this level, where the value of the probability density is locally
analyzed, there is no clue on whether there are or there are not
phase-related effects, such as interference traits; a
global view of the topology exhibited by the probability density is
necessary to get such an information, i.e., to observe the well-known
fringe structure.
This does not mean that it is not possible to get phase information
on equal footing; it is only that in quantum mechanics it is not
common to directly focus on the quantum phase, because it is
a quantum observable quantity (according to the conventional definition
of quantum observable), while the probability density is directly
related to intensities, experimentally accessible by accumulating
events (detections) for a given time.
As seen above, local phase information (not to be confused with global
phase factors) is encoded in the phase field $S$, defined by
(\ref{eq4}), which can be determined from the wave function.
In the recreation of the two-slit experiment here analyzed, the
time-evolution of this field along the transverse coordinate is readily
determined by substituting the wave function (\ref{eq1}) into
Eq.~(\ref{eq4}), and then assigning values for $x$ and $t$, as shown in
Fig.~\ref{fig3}(b).
For a better visualization, given the continuous increase of the
function as $|x|$ increases (at a given time), a $(-\pi,\pi]$
representation has been chosen (the vertical axis is given in units of
$\pi$ for clarity), which explains the sharp ``jumps'' from $\pi$ to
$-\pi$ around $|x| \approx 12.5$, for instance.
In any case, and neglecting the modulo operation to bound the value of
$S$ for clarity purposes, it is apparent that the topology exhibited by
the phase field resembles that of the probability density.
Or is it just that the evolution of the latter is strongly influenced
by the former?

\begin{figure*}[t]
\begin{center}
 \includegraphics[width=14cm]{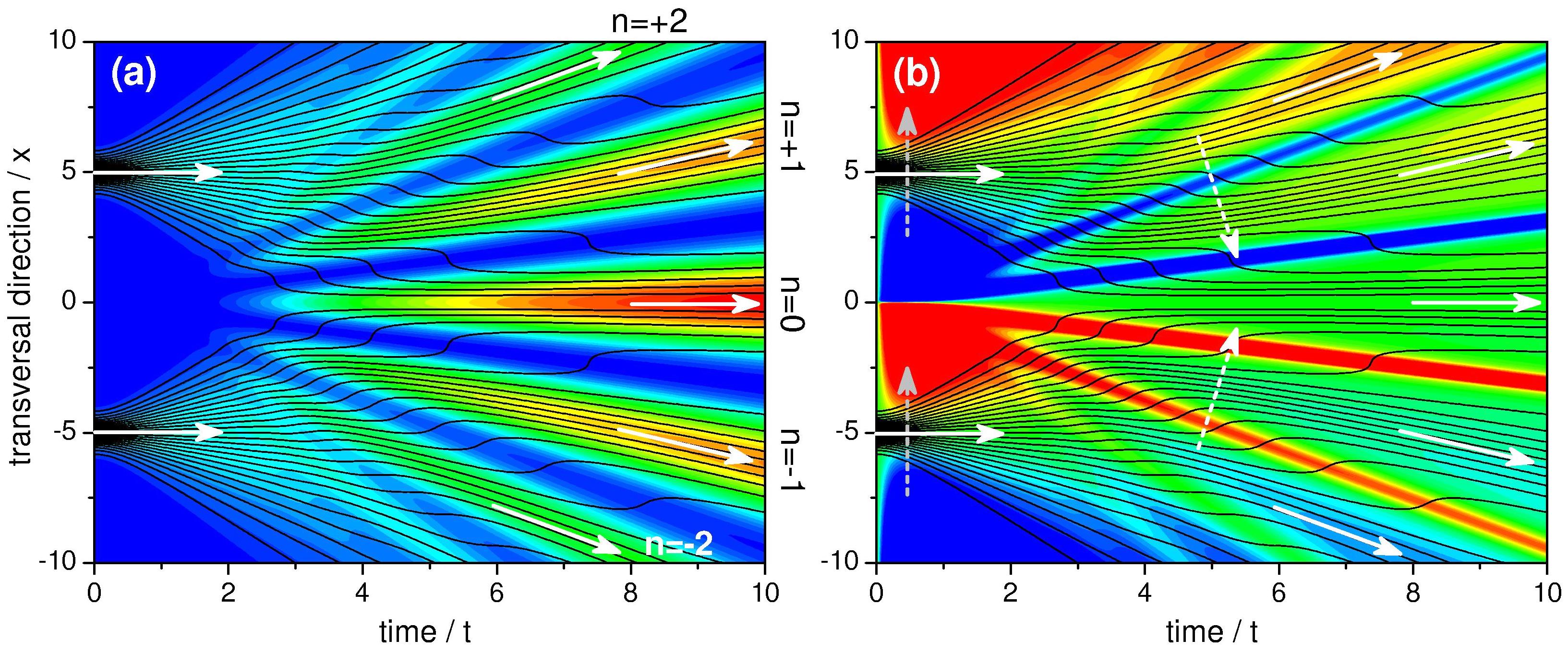}
 \caption{Numerical simulation illustrating the usual Bohmian
  interpretation of Young's two-slit experiment \cite{sanz:foundphys:2015}
  by means of Bohmian quantum trajectories (black solid lines).
  For the sake of clarity, these trajectories are superimposed on the
  contour plots of the probability density (a) and the velocity field
  (b).
  The trajectories represent the evolution in time (according to the
  Bohmian prescription, given by the guidance equation (\ref{eq8}))
  of a single
  element of quantum fluid, from any of the two slits to some final
  position where a scanning screen (detector) would be allocated.
  White arrows indicate the main (evolution) direction of the flow
  initially and asymptotically, which shows how from zero transversal
  flow the dynamics turns into a set of components traveling with
  different transverse speed.
  In part (b), the gray dashed arrows indicate the direction in which
  the velocity field increases at the early stages of the evolution,
  while the white dashed arrows show the trend motion followed by the
  different sub-ensembles of trajectories (downwards in the upper half
  of the graph; upwards in the lower half).}
 \label{fig4}
\end{center}
\end{figure*}

A quick answer can be given to such a question by invoking the so-called
Hamiltonian analogy, which links Fermat's and Huygen's optics principles
with Jacobi's mechanistic formulation of Maupertuis' principle, and was
reconsidered later on by Schr\"odinger to derive his wave equation by
merging de Broglie's relation with the Hamilton-Jacobi formulation of
classical mechanics.
Although it is often neglected (probably, always), this is actually in
conventional quantum mechanics, which provides a natural way to
understand the relationship between the probability density ($\rho$)
and the associated phase field ($S$).
Additionally, taking the Bohmian perspective (picture) of quantum
mechanics, the issue can be tackled even in a simple manner by
appealing to two distinctive elements of this view, namely the quantum
potential ($Q$) and the local velocity field ($v$).
The numerical evolution of these two fields is represented in
Figs.~\ref{fig3}(c) and (d), respectively.

In the case of the quantum potential [see Fig.~\ref{fig3}(c)], the near
field is essentially characterized by two inverted parabolas, which
correspond to the measure of the curvature of the wave function in
the neighborhood of the slits.
However, as time proceeds, $Q$ passes from displaying a rapidly
increasing ``dynamic barrier'' in the middle region between the two slits
(truncated red 3D surface representation in the figure, starting around
$t \approx 2$),
typical of the Fresnel regime, to an alternate structure of plateaus
and dips or ``canyons'' as we move further away towards the Fraunhofer
regime.
The central barrier in the Fresnel regime represents the effective meeting
of the two wave packets, that is, the point at which the single slit diffraction
process ends because the two waves start coalescing, giving rise to
the appearance of incipient interference features.
This process takes some time until a sort of ``equilibrium'' is reached,
typical of the Fraunhofer regime -- a regime where the overall shape of the
wave or, in this case, the quantum potential does not change, but spreads
linearly with time \cite{sanz:AJP:2012} (or, equivalent, the distance from
the two slits).

Comparing Figs.~\ref{fig3}(a) and (c), we note the direct
relationship between both $\rho$ and $Q$.
Actually, according to Eq.~(\ref{eq7}), the quantum potential is just
a sort of measure of the local curvature of the probability density,
so that the maxima of the probability density lie on the plateaus of
$Q$, while the minima do it on the dips.
Dynamically speaking, this means that nodes of the probability density
are associated with regions relatively unstable regions of the quantum
potential, while non-vanishing probability gives rise to regions of
relative stability.
This is precisely an important issue when designing quantum trajectory
methods \cite{wyatt-bk}, since regions with low densities generate
numerical instabilities, eventually translating into a source of inaccuracy
of the method.

The problem of interpreting the quantum potential as a potential
function is that it depends directly on the probability
density, thus providing us with redundant information.
That is, it does not matter whether we look at the quantum potential
or the probability density to understand that two initial ensembles of
trajectories leaving each slit (see discussion below) will eventually
split up into different sub-ensembles.
Nonetheless, there is an interesting difference worth mentioning
and also worth being taught to explain and understand interference at
a more intuitive level (even if one regards the quantum potential
itself as not intuitive at all).
The probability density describes the chance to find trajectories along some
particular directions, avoiding others.
It only gives us statistical information.
On the other hand, the quantum potential, appealing to a Newtonian-like view,
provides us with a certain kind of mechanistic information, specifying that the
trajectories evolve along those directions, because the corresponding quantum
force along them is negligible ($\nabla Q \approx 0$), while they avoid the
regions where this potential undergoes remarkable fast variations (intense
quantum forces), as can be seen in Fig.~\ref{fig3}(c).

So, probability density and quantum potential are face and tail of the same
coin.
Now, given that the quantum potential is directly connected to the probability
density, one may wonder if there is an alternative way to interpret and explain
the dynamics observed.
To answer this question, remember that $Q$ comes from the Laplacian (kinetic)
operator and therefore contains (and conveys) nonlocal dynamical information
unlike classical potential functions.
Consequently, this should manifest in some way.
This is the point where the quantum phase and its associated velocity field
come into play.
As seen in the previous section, the quantum phase is linked to the concept
of quantum flux, which in classical statistics is related to the evolution of
swarms of particles.
Let us then try this way and compute the velocity field arising from
the gradient of the phase field, which is plotted in Fig.~\ref{fig3}(d).
This field gives us an idea on the local variations undergone by the quantum
phase field [Fig.~\ref{fig3}(b)].
As time proceeds, within the Fraunhofer regime, the velocity field displays a
series of relative maxima and minima that lie on a nearly positive-increasing
slope (for a given time).
The maxima and minima have the shape of localized spikes (positive and
negative, respectively), which evolve precisely along the minima of the
probability density and denote regions where the velocity field is relatively
intense, i.e., trajectories will cross those regions with a rather high speed
(see Fig.~\ref{fig4}).
Regarding the overall slope mentioned before, it is interesting to note that
it develops from a full horizontal position at $t=0$ to an inclined (positive)
angle as time proceeds, from the Fresnel to the Fraunhofer regime
\cite{luis:AOP:2015}.
This behavior is governed by the presence of a strong correlation
between the dynamics exhibited by the two diffracted beams since
$t=0$.
It is the physical manifestation of what we call {\it quantum coherence},
which makes a clear distinction between a bare superposition of solutions
and its physical consequences: although at a practical level the probability
density looks the same independently of whether one first propagates one
wave packet and then the other (according to the usual {\it superposition
principle}), physically both wave packets must leave the slits at the same
time, because it is the phase field associated with both simultaneously
propagating what originates the particular dynamics exhibited by quantum
systems \cite{sanz:JPA:2008} (compared to classical systems).

As a result of the global nature of the quantum phase field and,
through it, the associated velocity field, the trajectories or
streamlines that the Bohmian picture renders have nothing to do with
those one typically consider in classical mechanics.
Actually, this nonlocal or global behavior is typical of ensembles of
streamlines mapping any kind of wave, regardless of its
nature \cite{sanz:AnnPhysPhoton:2010,milena:JRLR:2015}.
Specifically, in the case here considered, the trajectories tend to
move to regions with lower values of the modulus of the velocity field
(more stable, dynamically speaking), where they display what could be
regarded as a nearly classical-like uniform motion ($v \approx$
constant), as can be seen in Fig.~\ref{fig4}(b).
On the contrary, near sharp changes of the quantum phase, and
therefore large absolute values of the velocity field, the trajectories
undergo the effects of a sudden acceleration, which leads them from one
stability region to the neighboring one.
Notice that the changes of the velocity allow to explain in a better
way the motion displayed by the trajectories than the quantum
potential.
Specifically, the quantum potential does not explain, for instance, why
the motion should be symmetric with respect to $x=0$ --- apart from
extra symmetry arguments ---, while by looking at the sign of the
velocity field this effect can be perfectly understood [see arrows in
Fig.~\ref{fig4}(b)].

Regarding the physics described by the velocity field
(\ref{eq10}), it is worth stressing that its value coincides with the
value rendered after performing a ``weak measurement'' following the
usual or traditional view of the quantum theory, according to
Wiseman \cite{wiseman:NewJPhys:2007}.
This result was experimentally confirmed precisely from a realization
of Young's experiment with photons \cite{kocsis:Science:2011}.
Specifically, this experiment showed that the trajectories obtained
from Eq.~(\ref{eq10}) are compatible with the data rendered by weak
measurements of the average velocity associated with a swarm of
identically prepared photons.
These experiments have motivated the search of Bohmian trajectories in
different systems \cite{braverman:PRL:2013,schleich:PRA:2013}.
Of course, this does not mean that we know the specific path chosen by
a particle in Young's renowned two-slit experiment \cite{shamos-bk},
but only that the average flow of a large number of such particles,
described by the same single-particle wave function, is laminar, which
cannot easily be inferred a priori from other pictures, although it is
a perfectly reachable result through the flux operator
\cite{sanz:JPA:2008}.


\section{Concluding remarks}
\label{sec5}

In the centennial of David Bohm's birth anniversary, when his
``suggested interpretation'' of quantum phenomena is turning 65, it
is probably time to start thinking in more natural terms this approach,
showing and teaching it from a different perspective, closer to the
conception Bell had about it \cite{bell-bk}.
This has been the goal of this work, tackled through the two questions
posed in the introductory part.
Accordingly, an alternative view of Bohmian mechanics has been
presented, as a complementary picture of quantum mechanics on equal
footing with other more widely known and used pictures.
To this end, the work has been split up into two independent but
complementary blocks, one dealing with a brief historical perspective
and another one about specific aspects of Bohmian mechanics and its
application to the description of quantum interference.

Combining both blocks, we find that there is no need for appealing to
the concept of {\it hidden variable} when talking about Bohmian mechanics at
present.
Rather, this is just another alternative and complementary picture of
quantum mechanics, which provides us with hydrodynamic-like information
of quantum systems that does not contradict the kind of information
rendered by other pictures, as mentioned above.
Actually, because we have at hand this kind of information in terms of
trajectories, it is possible to apply techniques typical of classical
mechanics and hydrodynamics for its analysis, thus unraveling
interesting properties and determining alternative descriptions
and explanations for the evolution of quantum systems, particularly
useful to understand the role played in them by the quantum phase and
quantum correlations (entanglement) \cite{note2new}.
This is illustrated here by means of a simple realization of Young's
two-slit experiment, which combines both analytical and numerical
aspects feasible to be taught and developed in a standard elementary
course on quantum mechanics.
The fact that such evolution can be followed by means of trajectories
does not contradict at all our understanding of the quantum theory, but
is in compliance with it provided we do not assign any reality to such
trajectories (or a direct link between them and the actual motion
displayed by a real quantum particle, as hidden-variable models do).
Notice there is no experimental evidence that allows us to
establish it -- recent experiments only confirm that the average
flux is compatible with these trajectories \cite{kocsis:Science:2011}.


\section*{Acknowledgments}

This work intends to be a posthumous tribute to David Bohm on the
occasion of the centennial of his birthday.
He was willing to look at quantum mechanics with different eyes in
times where all eyes looked on the same direction.
And, in so doing, he and his works served as an inspiration to
forthcoming generations of physicists.
The author is grateful to Profs.\ Basil Hiley and Jos\'e Luis S\'anchez
G\'omez for enjoyable hours of conversations on different aspects
concerning Bohmian mechanics, hidden variables and the foundations
of the quantum theory.
The author would also like to acknowledge fruitful feedback from
Profs.\ Hrvoje Nicoli\'c and Chris Dewdney.

This work has benefitted from financial support from the Spanish MINECO
(Grant No.\ FIS2016-76110-P).



\begin{thebibliography}{108}
\bibitem{note1}
Following Schiff \cite{schiff-bk} (see p.~171), here the term
``picture'' will also be used instead of ``representation'', leaving
the latter ``for designation of the choice of axes in Hilbert space
or, equivalently, the choice of the complete orthonormal set of
functions, with respect to which the states and dynamical variables
are specified.''
\bibitem{note3}
 Throughout this work, the concept of quantum particle or system is used to
 denote the degrees of freedom of interest necessary to characterize and study
 a real physical system.
 If such degrees of freedom are translational, they
 can be associated without loss of generality with the particle itself,
 since they account for its center of mass, as in classical mechanics.
 However, this is a particular case, since they may also
 describe vibrations, rotations, etc., and although Schrodinger's
 equation and its Bohmian reformulation are still valid in these cases,
 they no longer refer to particles in the usual sense.
\end{thebibliography}

\end{document}